\documentclass[fleqn,usenatbib]{mnras}

\usepackage{newtxtext,newtxmath}

\usepackage[T1]{fontenc}

\DeclareRobustCommand{\VAN}[3]{#2}
\let\VANthebibliography\thebibliography
\def\thebibliography{\DeclareRobustCommand{\VAN}[3]{##3}\VANthebibliography}

\usepackage{graphicx}
\usepackage{amsmath}

\title[Energy Source for Late-Time Kilonovae]{Radioactive Decay of Specific Heavy Elements as an Energy Source for Late-Time Kilonovae and Potential JWST Observations}

\author[M.-H. Chen and E.-W. Liang]{
Meng-Hua Chen$^{1,2}$ and En-Wei Liang$^{2}$\thanks{E-mail: lew@gxu.edu.cn}
\\
$^{1}$Kavli Institute for Astronomy and Astrophysics, Peking University, Beijing 100871, China\\
$^{2}$Guangxi Key Laboratory for Relativistic Astrophysics, School of Physical Science and Technology, Guangxi University, Nanning 530004, China\\
}

\date{Accepted XXX. Received YYY; in original form ZZZ}

\pubyear{2023}

\begin{document}
\label{firstpage}
\pagerange{\pageref{firstpage}--\pageref{lastpage}}
\maketitle

\begin{abstract}
Revealing the temporal evolution of individual heavy elements synthesized in the merger ejecta from binary neutron star mergers not only improves our understanding of the origin of heavy elements beyond iron but also clarifies the energy sources of kilonovae. In this work, we present a comprehensive analysis of the temporal evolution of the energy fraction of each nuclide based on the $r$-process nucleosynthesis simulations. The heavy elements dominating the kilonova emission within $\sim100$~days are identified, including $^{127}$Sb, $^{128}$Sb, $^{129}$Sb, $^{130}$Sb, $^{129}$Te, $^{132}$I, $^{222}$Rn, $^{223}$Ra, $^{224}$Ra, and $^{225}$Ac.
It is found that the late-time kilonova light curve ($t\gtrsim20$~days) is highly sensitive to the presence of the heavy element $^{225}$Ac (with a half-life of 10.0~days). Our analysis shows that the James Webb Space Telescope (JWST), with its high sensitivity in the near-infrared band, is a powerful instrument for the identification of these specific heavy elements. 
\end{abstract}


\section{Introduction}
\label{intro}

The mergers of binary neutron stars or neutron star-black hole binaries have long been speculated to be a promising site for the production of heavy elements beyond iron via the $r$-process nucleosynthesis in the universe \citep{Lattimer1974}.
The radioactive decay of newly synthesized $r$-process nuclei in the merger ejecta produces an electromagnetic transient in the ultraviolet/optical/near-infrared wavelengths, known as a kilonova (\citealp{Li1998,Metzger2010,Korobkin2012,Barnes2013,Kasen2013,Barnes2016}; see \citealp{Metzger2019} for a recent review). The radioactive decay of unstable nuclei may also produces a gamma-ray transient accompanying the kilonova \citep{Hotokezaka2016,Li2019,Chen2021,Chen2022} and a MeV neutrino flash \citep{Chen2023}. The solid detection of the kilonova AT2017gfo, associated with gravitational wave event GW170817, provides strong evidence to support this speculation \citep{Abbott2017a,Abbott2017b,Abbott2017c}.

The luminosity and color evolution of AT2017gfo are well consistent with the radioactively powered kilonova model, suggesting that such mergers  could be the major sites for $r$-process nucleosynthesis \citep{Cowperthwaite2017,Drout2017,Evans2017,Kasliwal2017,Nicholl2017,Pian2017,Smartt2017,Tanvir2017}. The analysis of the kilonova AT2017gfo reveals that the ejected material has two distinct components: an early ``blue'' kilonova and a later ``red'' kilonova \citep{Kasen2017,Kilpatrick2017,Nicholl2017,Villar2017}. The blue kilonova is interpreted as a signature of light $r$-process nuclei with atomic mass number $A\lesssim140$, while the red kilonova is commonly believed to be powered by a lanthanide-rich ejecta with high opacity ($\kappa\sim10$~cm$^2$~g$^{-1}$).
It is worth noting that all of these studies have considered the radioactive decay of heavy elements as the energy source powering the kilonova transient. However, the late-time kilonova emission may also be attributed to the energy injection from a long-lived remnant neutron star \citep{Yu2018,Ren2019}.

Determining the nuclide composition of the merger ejecta, especially the nulides located around the $r$-process peaks and the lanthanide elements, is crucial for understanding the origin of heavy elements and the energy sources of kilonovae.
Several research groups have been investigating the connection between the heavy elements and kilonova light curves \citep{Metzger2010,Lippuner2015,Wanajo2018,Watson2019,Wu2019,Kasliwal2022}. \cite{Metzger2010} were pioneers in estimating the radioactive heating rates and predicting kilonova light curves resulting from neutron star mergers by using $r$-process nucleosynthesis simulations. \cite{Lippuner2015} conducted a parameter study of $r$-process calculations and found that the ejected materials are lanthanide-free for high electron fractions ($Y_{\rm e}\gtrsim$ 0.25). \cite{Zhu2018} included the contribution from fission fragments to calculate the kilonova light curves from neutron star mergers. \cite{Wanajo2018} focused on calculating the contributions from individual $r$-process elements and found that the heavy elements around the second $r$-process peaks dominate the radioactive heating rates. The impact of nuclear physics uncertainties on $r$-process abundances patterns and kilonova emission was investigated by \cite{Barnes2021}. Moreover, \cite{Wu2019} suggested that heavy elements can be distinguished from kilonova light curves because the radioactive power is often dominated by a few nuclei, which can result in identifiable features in the light curves, such as bump-like or exponential decay-like features in the bolometric light curve. This distinctiveness could potentially be utilized to infer the yield of heavy elements in future kilonova observations \citep{Wu2019,Kasliwal2022}.

Given its high sensitivity in the near-infrared band, the James Webb Space Telescope (JWST) is expected to be a powerful instrument for follow-up observations and providing a comprehensive view of the temporal and spectral evolution of kilonovae from early to late-time phase \citep{Bartos2016}. Such observations offer an opportunity to reveal the production of individual heavy elements.
In this paper, we study this issue by analyzing the temporal evolution of energy fraction of each nuclide based on detailed $r$-process nucleosynthesis simulations. We investigate the heavy nuclei that dominate the kilonova emission within $\sim100$~days, and find out distinguished features in their bolometric light curves. We further explore whether these feature are detectable using the JWST telescope.
The paper is organized as follows. In Section~\ref{method}, we describe the methodology used to calculate the kilonova light curves. Our results are presented in Section~\ref{result}. Conclusions and discussions are given in Section~\ref{conclusion}.

\section{Methods}
\label{method}

\subsection{Nucleosynthesis}

We employ the nuclear reaction network code SkyNet\citep{Lippuner2015,Lippuner2017} to perform simulations of $r$-process nucleosynthesis. SkyNet evolves the abundances of 7843 nuclide species, ranging from free neutrons (atomic number $Z=0$) and protons ($Z=1$) to $^{337}$Cn ($Z=112$), and includes over 140 000 nuclear reactions. 
The nuclear reaction rates used in SkyNet are taken from the JINA REACLIB database \citep{Cyburt2010}. In our calculations, we use the latest nuclear database, including nuclear mass data from AME2020 \citep{Wang2021} and radioactive decay data from NUBASE2020 \citep{Kondev2021}.
For nuclide species without experimental data, we use the theoretical data derived from the Finite-Range Droplet Model \citep{Moller2016}. Neutron capture rates are calculated using the corresponding nuclear mass data through the nuclear reaction code TALYS \citep{Goriely2008}. Nuclear fission processes are calculated using fission barriers from \cite{Moller2015} and double Gaussian fission fragment distributions from \cite{Kodama1975}.

In our $r$-process nucleosynthesis simulations, we use the parameterized trajectory provided by \cite{Lippuner2015}. The initial conditions are consistent with the results of numerical relativity simulations given by \cite{Radice2018}.
We adopt the results of $(1.40 + 1.40)M_{\odot}$ binary system, which is simulated using three distinct equations of state: BHBlp\_M140140\_LK, DD2\_M140140\_LK, and LS220\_M140140\_LK. The initial electron fractions for these binary system are $Y_{\rm e}=$ 0.15, 0.17, 0.14, and the initial specific entropies are $s=$ 18, 22, 16~$k_{\rm B}$/baryon. Based on the numerical relativity simulations conducted by \cite{Radice2018}, the expansion timescale $\tau$ can be estimated as $\tau \approx cRe / 3v_R$, where $e \simeq 2.718$ is the Euler's number and $v_R$ corresponds to the velocity measured on a sphere with a coordinate radius of $R = 300 G/c^2 M_{\odot} \simeq 443$~km. Thus the expansion timescales of these binary system are 7.87~ms, 6.08~ms, and 7.87~ms, repectively.
Additionally, we consider the result of the unequal mass binary $(1.40 + 1.20)M_{\odot}$, labeled LS220\_M140120\_M0. This system has an initial electron fraction of $Y_{\rm e}=0.18$, an initial entropy of $s=14~k_{\rm B}$/baryon$^{-1}$, and an expansion timescale of $\tau=8.92$ ms).
In the calculations of radioactive power, we use the recent nuclear data from the Evaluated Nuclear Data File library (EDNF/B-VIII.0, \citealp{Brown2018}).
We note that we do not include the extra radioactive energy resulting from fission fragments since the EDNF library contains very few radiation data for fission processes. \cite{Zhu2018} have identified the specific fission reactions responsible for the radioactive power by using a double Gaussian fragment distribution from \cite{Kodama1975}.

\subsection{Radioactive Heating Rate}

The effective heating rate of merger ejecta is obtained by summing the contributions of all decay products, including $\alpha$-particles, $\beta$-particles, and $\gamma$-rays, i.e.,
\begin{equation}
\dot{Q}(t)=\sum_{\rm p} f_{\rm p}(t)\dot{q}_{\rm p}(t),~~~{\rm p}\in\{\alpha, \beta, \gamma \},
\end{equation}
where $f_{\rm p}(t)$ is the thermalization efficiency.
The radioactive heating rate $\dot{q}_{\rm p}(t)$ is given by the sum of the energy released by all nuclides in the ejecta,
\begin{equation}
\dot{q}_{\rm p}(t)= N_{\rm A} \sum_i \lambda_i Y_i(t) E_{\rm p},~~~{\rm p}\in\{\alpha, \beta, \gamma \}
\end{equation}
where $N_{\rm A}$ is the Avogadro's number, $\lambda_i$ is the decay rate of the $i$th nuclide, $Y_i(t)$ is the corresponding abundance, and $E_{\rm p}$ is the energy released by individual nuclear reaction.
The decay rate is given by $\lambda_i=\ln2/T_{1/2,i}$ with $T_{1/2,i}$ being the half-life of the $i$th nuclide.
The thermalization efficiencies for each decay product are estimated by following the analytic methods presented in \cite{Kasen2019}.\\
For $\beta$-particles,
\begin{equation}
f_{\beta}(t)=(1+t/t_{\beta})^{-1},
\end{equation}
where $t_{\beta}$ is the characteristic timescale at which thermalization starts to become inefficient,
\begin{equation}
t_{\beta}=12.9~\left[\frac{M_{\rm ej}}{0.01M_{\odot}}\right]^{2/3} \left[\frac{v_{\rm ej}}{0.2c}\right]^{-2}\zeta^{2/3}~{\rm days},
\end{equation}
where $M_{\rm ej}$ is the total ejecta mass, $v_{\rm ej}$ is the characteristic ejecta velocity, and $\zeta$ is a constant defined to close to unity (we adopt $\zeta=1$). \\
For $\alpha$-particles,
\begin{equation}
f_{\alpha}(t)=(1+t/t_{\alpha})^{-1.5},
\end{equation}
where the thermalization timescale of $\alpha$-particles is roughly $t_{\alpha}\approx3 t_{\beta}$.\\
For $\gamma$-rays,
\begin{equation}
f_{\gamma}(t)=1-\exp\left[-\frac{t_{\gamma}^2}{t^2}\right],
\end{equation}
where the thermalization timescale $t_{\gamma}$ is given by
\begin{equation}
t_{\gamma}=0.3~\left[\frac{M_{\rm ej}}{0.01M_{\odot}}\right]^{1/2}~\left[\frac{v_{\rm ej}}{0.2c}\right]^{-1}~{\rm days}.
\end{equation}
In this paper, we adopt an  ejecta mass of $M_{\rm ej}=0.04M_{\odot}$ and an ejecta velocity of $v_{\rm ej}=0.15c$. These values are consistent with the estimations derived from the red kilonova component observed in AT2017gfo, as reported by \cite{Kasen2017}.

\subsection{Kilonova Model}

We assume a spherical merger ejecta as that adopted by \cite{Metzger2019} . The merger ejecta is divided into $N$ ($N=100$) layers and each layer evolves nearly independently (see also \citealp{Chen2021,Chen2022}).
The expansion velocity of the $n$th layer is given as $v_n=v_0 + n\Delta v$, where $n=1, 2,\ldots, 100$, $\Delta v=0.001c$, and $v_0$ is a free parameter.
The radius of the $n$th layer at time $t$ is given by $R_n=v_n t$ and its mass can be estimated by $m_n=\int_{R_{n}}^{R_{n+1}}4\pi \rho r^2dr$.
The merger ejecta is assumed to have a power-law density profile \citep{Nagakura2014}
\begin{equation}
\rho(v_n, t)=\rho_0(t)\left(\frac{v_n}{v_0}\right)^{-3},
\end{equation}
where $\rho_0(t)$ is the density of the innermost layer.
For a given ejecta mass $M_{\rm ej}$, $\rho_0(t)$ can be derived from
\begin{equation}
M_{\rm ej} = \int_{R_{\rm 1}}^{R_{\rm N}}4\pi\rho(v_n,t)r^2dr.
\end{equation}
The characteristic ejecta velocity can be estimated as
\begin{equation}
v_{\rm ej}=(2 E_{\rm kin}/M_{\rm ej})^{1/2},
\end{equation}
where $E_{\rm kin}=\sum_n m_n v_n^2/2$ is the sum of the kinetic energy of all layers. The velocity of the innermost layer, $v_0$, is determined by the characteristic ejecta velocity $v_{\rm ej}$ through the equation:
\begin{equation}
\frac{1}{2}M_{\rm ej}v_{\rm ej}^2=\sum_n \frac{1}{2} m_n (v_0 + n\Delta v)^2.
\end{equation}

Following \cite{Metzger2019}, the thermal energy of each layer evolves according to
\begin{equation}
\frac{d E_n}{d t}=-\frac{E_n}{R_n}\frac{dR_n}{dt}-L_n+\dot{Q}(t)m_n,
\end{equation}
where $E_n$ is the internal energy of the $n$th layer, $L_n$ is the radiation luminosity, and $\dot{Q}(t)$ is the $r$-process heating rate.
The radiation luminosity of each layer can be estimated by
\begin{equation}
L_n=\frac{E_n}{t_{{\rm lc}, n}+t_{{\rm d}, n}},
\end{equation}
where $t_{{\rm lc},n}=v_n t / c$ is the light crossing time and $t_{{\rm d}, n}=\tau_n v_n t /c$ is the photon diffusion timescale.
The optical depth from the $n$th layer to the observer $\tau_n$ is given by
\begin{equation}
\tau_n=\int_{R_n}^{\infty} \rho(v_n,t)\kappa dr,
\end{equation}
where $\kappa$ is the ejecta opacity. We set $\kappa=10$~cm$^2$~g$^{-1}$ in our calculation, which is a reasonable approximation for a lanthanide-rich ejecta \citep{Tanaka2018}.
The total bolometric luminosity of the merger ejecta can be obtained by summarizing the contributions of all layers,
\begin{equation}
L_{\rm bol}=\sum_n L_n.
\end{equation}
The effective temperature of kilonova emission can be determined by
\begin{equation}
T_{\rm eff}=\left(\frac{L_{\rm bol}}{4\pi\sigma_{\rm SB}R_{\rm ph}^2}\right)^{1/4},
\end{equation}
where $\sigma_{\rm SB}$ is the Stephan-Boltzmann constant.
The radius of the photosphere $R_{\rm ph}$ is defined by the layer at which the optical depth is unity, i.e.,
\begin{equation}
\tau_{\rm ph}=\int_{R_{\rm ph}}^{\infty}\rho(v_n,t)\kappa dr=1.
\end{equation}
The flux density of the kilonova emission at frequency $\nu$ can be given by
\begin{equation}
F_\nu(t)=\frac{2\pi h\nu^3}{c^2}\frac{1}{{\rm exp}({h\nu/k T_{\rm eff}})-1}\frac{R_{\rm ph}^2}{D_L^2},
\end{equation}
where $h$ is the Planck constant, $k$ is the Boltzmann constant, and $D_L$ is the luminosity distance.
Then, the monochromatic AB magnitude is given by
\begin{equation}
M_\nu(t)=-2.5\log_{10}\left(\frac{F_\nu(t)}{3631~{\rm Jy}}\right).
\end{equation}

\section{Results}
\label{result}

\begin{figure}
\centering
\includegraphics[width=0.45\textwidth]{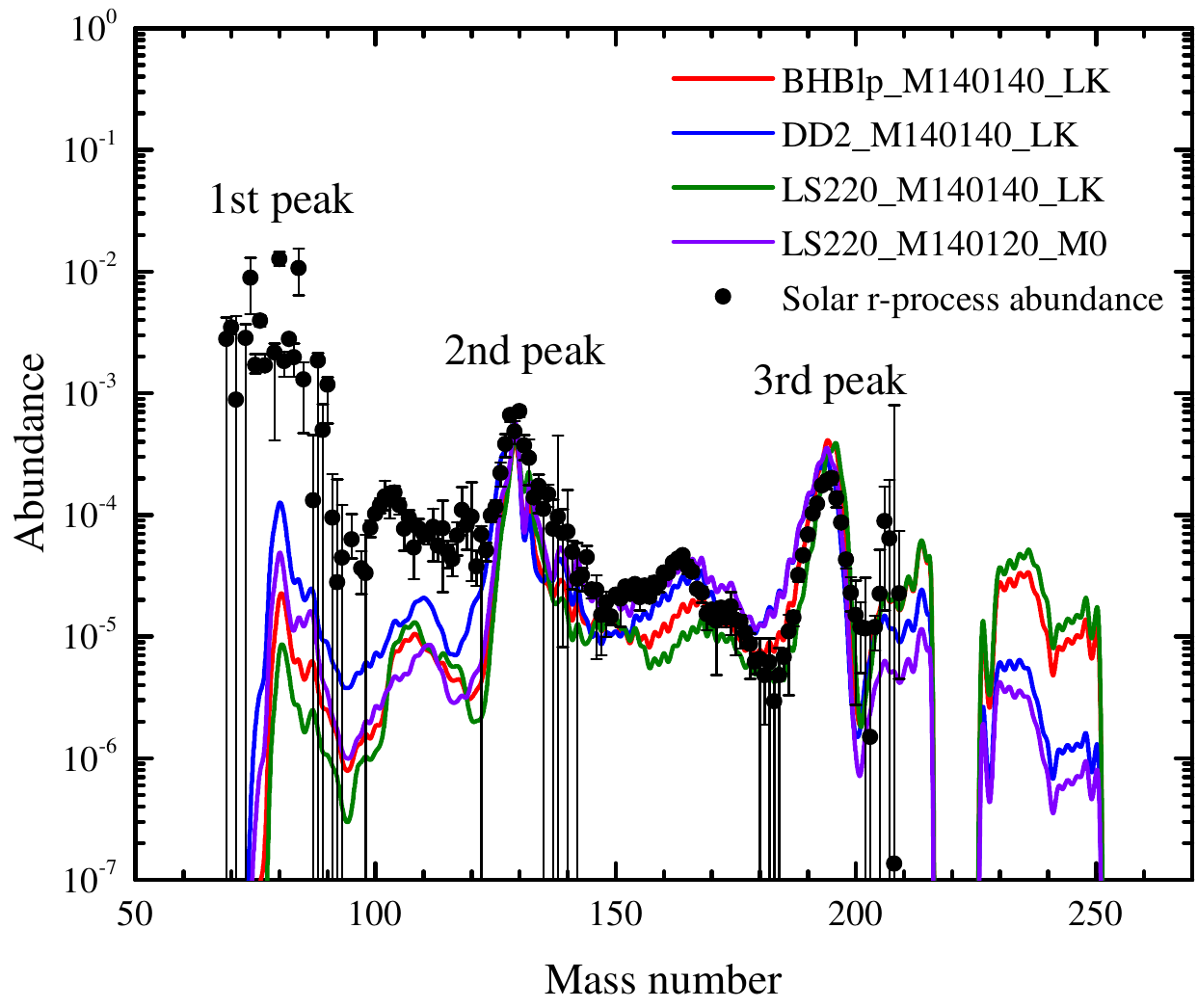}
\caption{Abundance patterns for $r$-process nucleosynthesis simulations. The initial conditions are consistent with the results of numerical relativity simulations \citep{Radice2018}. }
\label{Abundance}
\end{figure}

\begin{figure}
\centering
\includegraphics[width=0.45\textwidth]{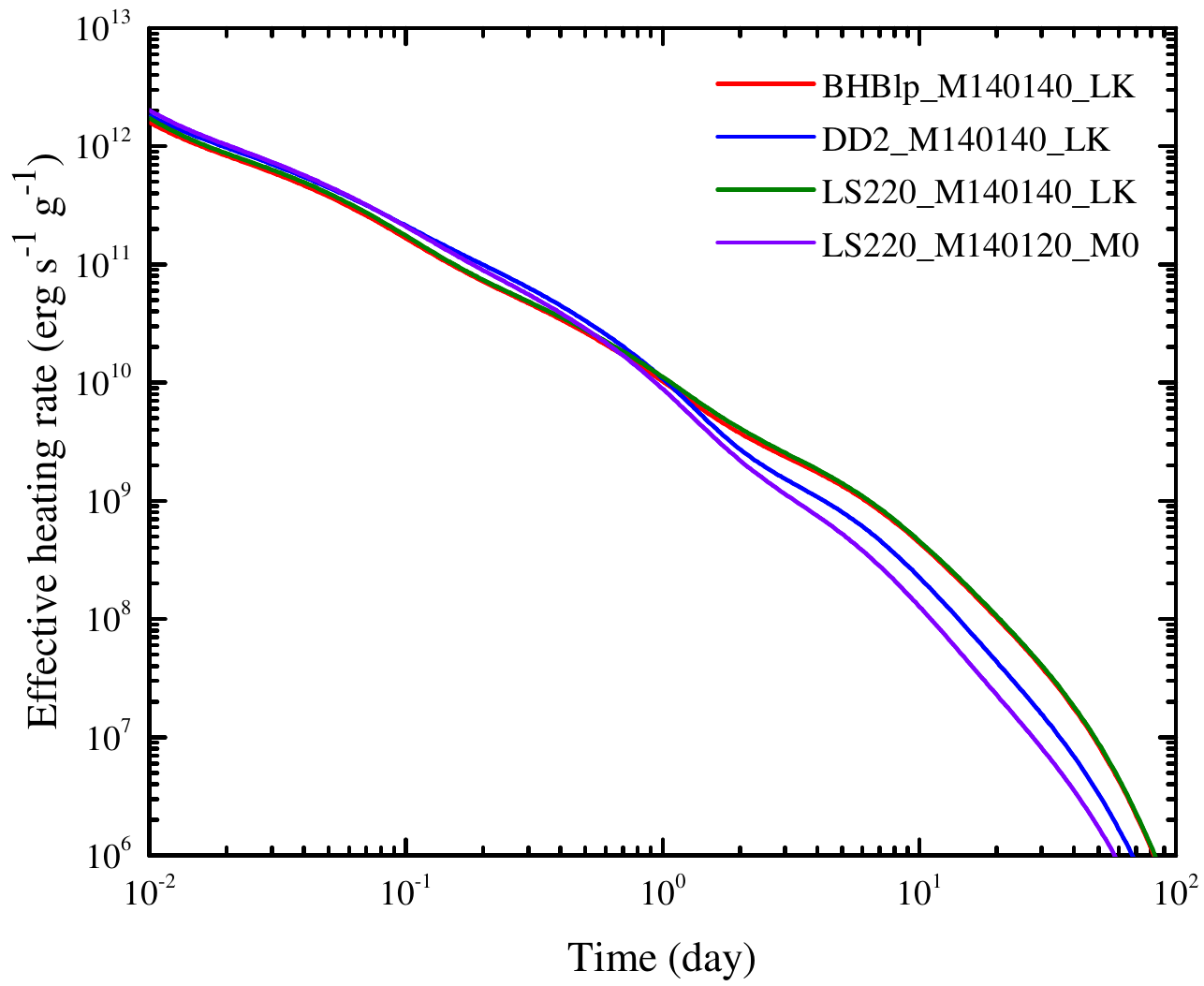}
\caption{Effective heating rates produced by the radioactive decay of $r$-process elements.}
\label{Heating}
\end{figure}

\begin{figure}
\centering
\includegraphics[width=0.45\textwidth]{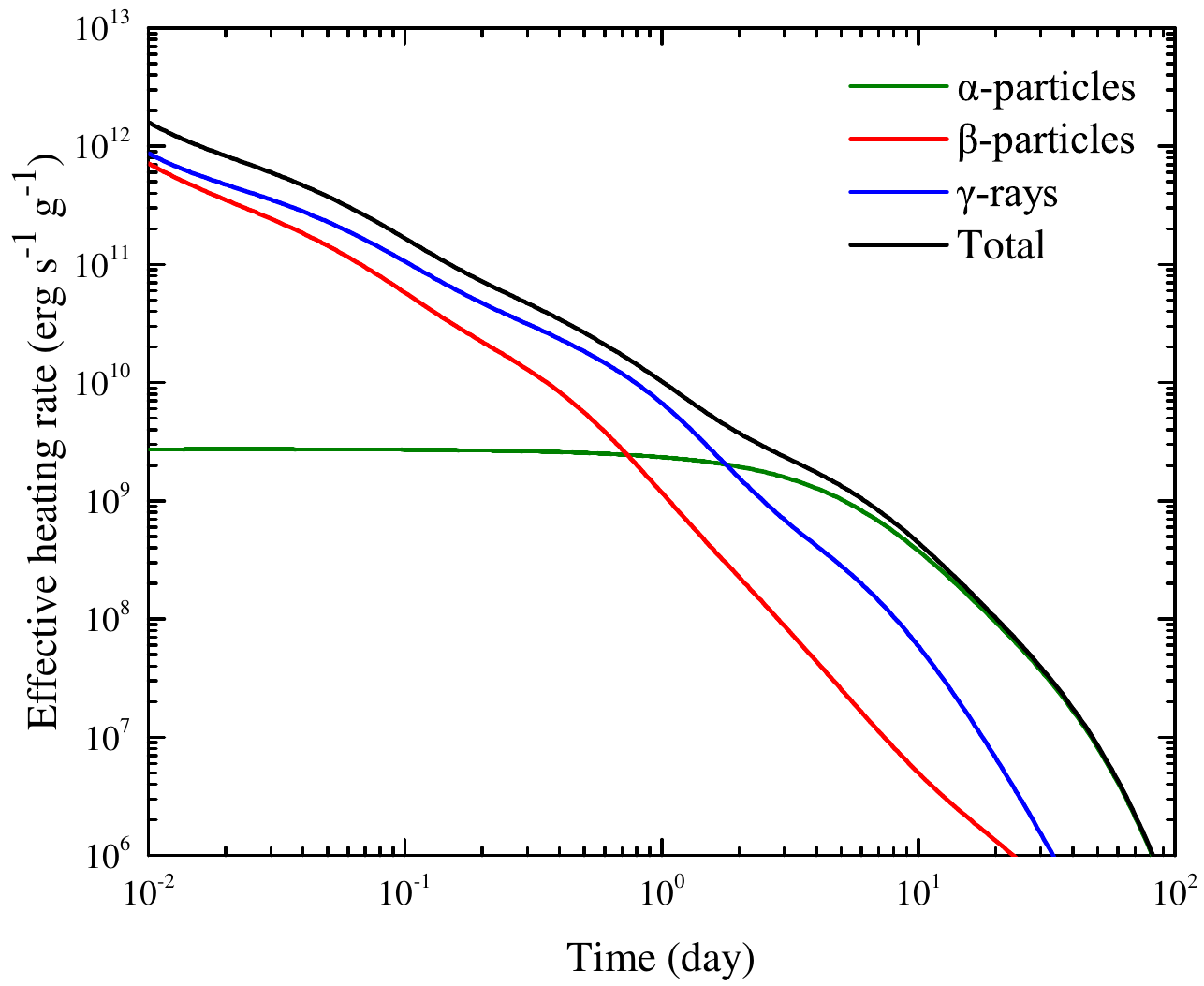}
\caption{The radioactive energy generated by $\alpha$-particles, $\beta$-particles, and $\gamma$-rays, respectively.}
\label{Fraction}
\end{figure}

Figure~\ref{Abundance} shows the abundance patterns obtained from $r$-process nucleosynthesis simulations. It is found that the merger ejecta is sufficiently neutron rich, enabling the production of a wide range of heavy elements from the second to third $r$-process peak. The $r$-process simulations calculated with various astrophysical trajectories all produce a rather broad distribution of heavy elements, and the overall abundance patterns are in agreement with the observed solar $r$-process abundances, particularly for atomic mass numbers $A\ge120$.

Based on the $r$-process nucleosynthesis simulations, we calculate the effective heating rate produced by the radioactive decay of newly synthesized heavy elements in neutron star mergers, as shown in Figure~\ref{Heating}. It is shown that the discrepancies in effective heating rates across different equations of state are modest. Figure~\ref{Fraction} shows the decay products responsible for effective heating rate. The total heating rate mainly comes from the radioactive energy of $\alpha$-particles, $\beta$-particles, and $\gamma$-rays. It is shown that $\beta$-particles and $\gamma$-rays play a crucial role in the heating rate during the early stage ($t\lesssim10$~days), while $\alpha$-particles dominate the nuclear heating at later times. This is caused by the fact that the lifetimes of $\alpha$-decay nuclei are typically longer than those of $\beta$-decay nuclei \citep{Hotokezaka2020}.

\begin{figure}
\centering
\includegraphics[width=0.45\textwidth]{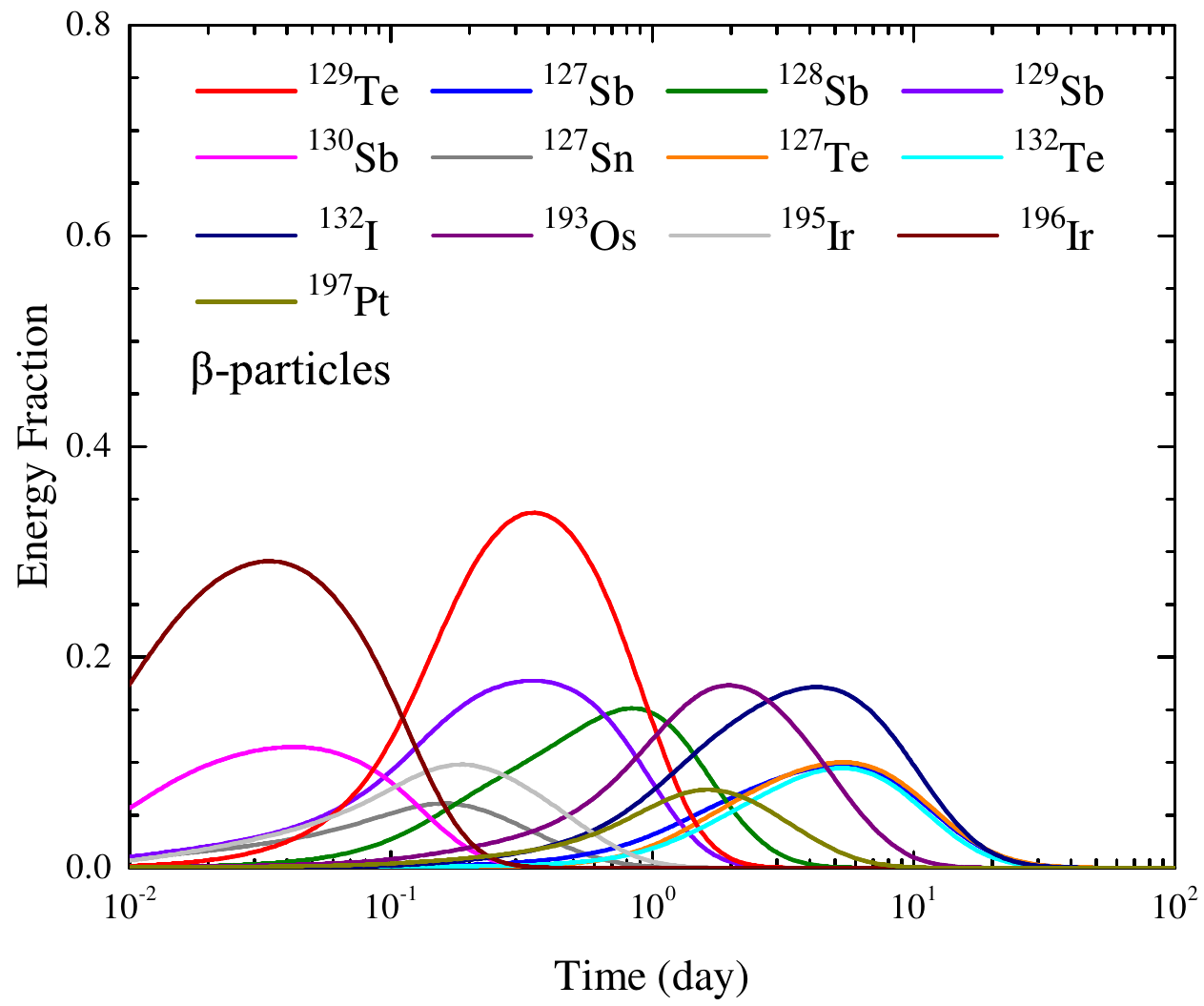}\\
\includegraphics[width=0.45\textwidth]{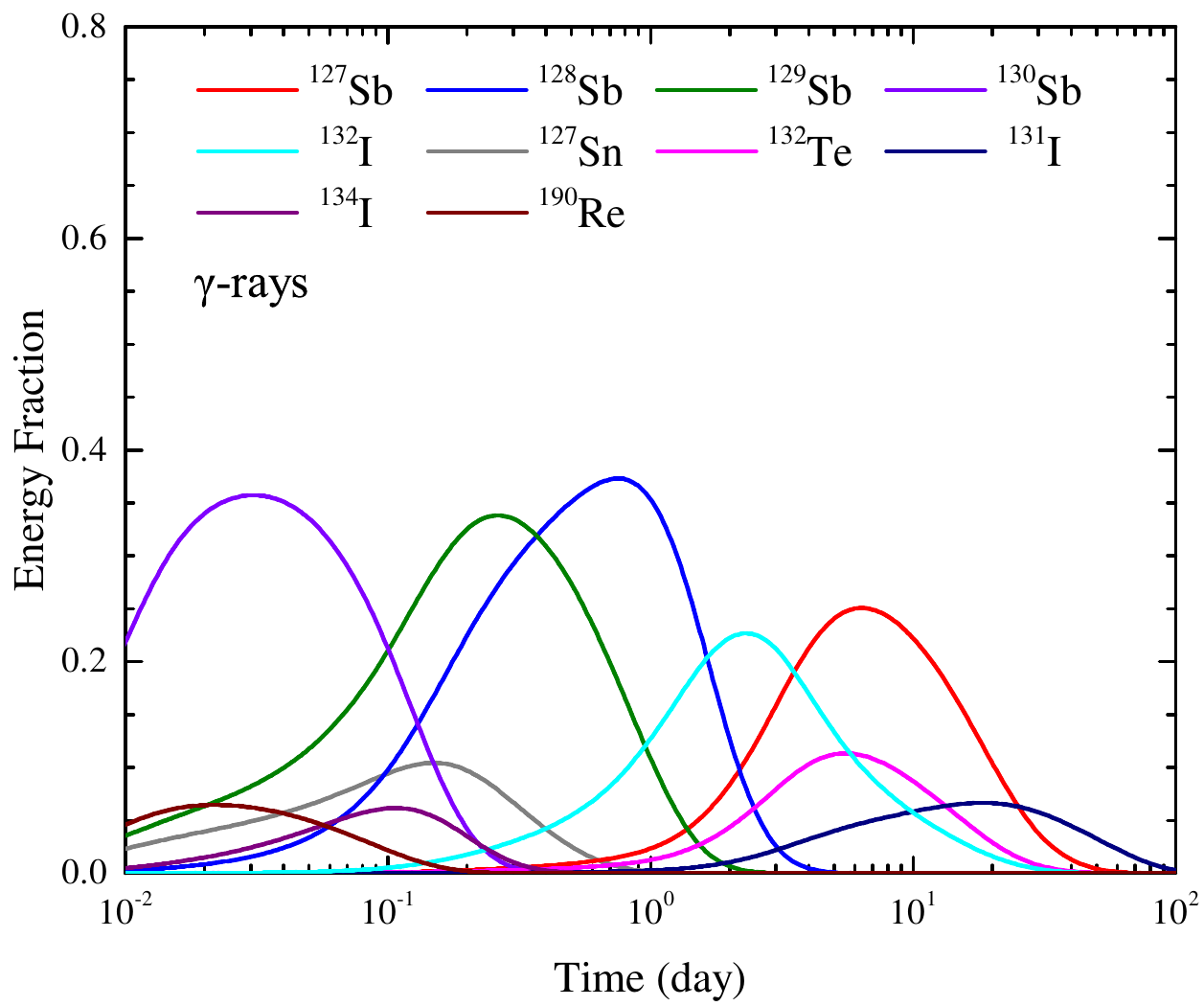}\\
\includegraphics[width=0.45\textwidth]{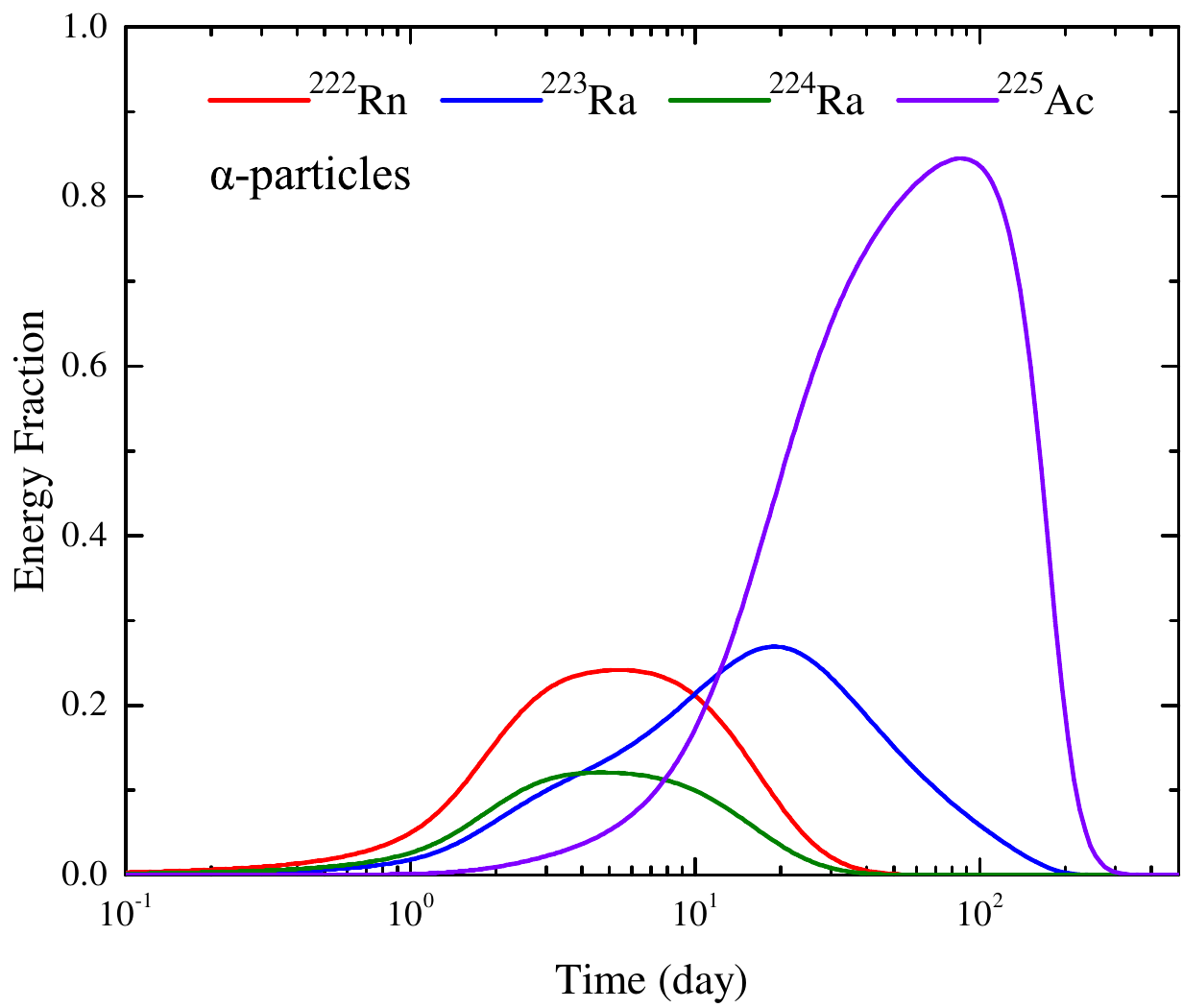}
\caption{The energy fraction of the dominant nuclides that contribute significantly to the total energy generation within 100~days for $\beta$-particles (top panel), $\gamma$-rays (middle panel), and $\alpha$-particles (bottom panel), respectively. Here we take BHBlp\_M140140\_LK as a representative case for our analysis.}
\label{nuclide}
\end{figure}

To investigate the nuclides that make dominant contribution to the radioactive heating rate, we calculate the energy generation of each nuclide. Given the insignificant discrepancies in radioactive heating rates when using different astrophysical trajectories, we take BHBlp\_M140140\_LK as the representative case for our analysis. In the top panel of Figure~\ref{nuclide}, we show the energy fraction of the dominant nuclides that contribute significantly to the total energy generation of $\beta$-particles between 0.01~day and 100~days. It is clearly shown that radioactive heating rate is powered by several dominant nuclides with different lifetimes. Each dominant nuclide has a characteristic time when it plays a crucial role in the radioactive heating of the ejecta. For example, the radioactive decay of $^{129}$Te generates over one-third of the total $\beta$-particles energy at $\sim0.35$~day. Therefore, the kilonova emission at $\sim0.35$~day may sensitive to $^{129}$Te since it produces the most significant energy at that moment. This suggests that it may be possible to distinguish the dominant nuclides from kilonova light curves when the contributions of other nuclei are relatively smaller or subdominant. It should be noted that the contributing nuclides may depend on the expansion models and nuclear ingredients adopted.

The dominant nuclides for the total $\gamma$-rays energy are also identified, as shown in the middle panel of Figure~\ref{nuclide}. It is found that the top contributing nuclides of the total $\gamma$-rays energy are almost common among all four models studied, including $^{127}$Sb, $^{128}$Sb, $^{129}$Sb, $^{130}$Sb, and $^{132}$I. These top five nuclides contribute to more than 20\% of the total $\gamma$-rays energy at their characteristic time. For nuclides with characteristic time $t_{\rm c}\gtrsim1$~day, their $\gamma$-rays may escape from ejecta without thermalization and form a $\gamma$-ray transient, since merger ejecta becomes transparent to $\gamma$-rays at about a day after merger. Hence, the radioactive nuclei of $^{128}$Sb, $^{129}$Sb, and $^{130}$Sb with characteristic times $t_{\rm c}\lesssim1$~day could be distinguished from kilonova light curves, but $^{127}$Sb and $^{132}$I with characteristic times $t_{\rm c}\gtrsim1$~day could be directly identified by MeV gamma-ray detectors. It should be noted that the identification of gamma-ray lines produced by radioactive decay of heavy elements may be possible only for quite rare, nearby events \citep{Hotokezaka2016,Li2019,Chen2021,Chen2022}.

The energy fraction of the dominant $\alpha$-decay chains that contribute significantly to the total heating rate is shown in the bottom panel of Figure~\ref{nuclide}. These $\alpha$-decay chains are responsible for the late-time heating rate due to they liberate more energy per decay and thermalize with greater efficiency than $\beta$-decay products. For instance, the  $\alpha$-decay $^{225}$Ac (with an $\alpha$-decay half-life of $10.0$~days, originating from the $\beta$-decay of $^{225}$Ra with a $\beta$-decay half-life of $14.9$~days) releases 19.32~MeV, which is significantly higher than the energy released by the $\beta$-decay process. Since no other radioactive nuclei can release a comparable amount of energy within this timescale, $^{225}$Ac is the dominant contributor to the late-time kilonova emission. Furthermore, the initial conditions for $r$-process calculations provided by numerical relativity simulations \citep{Radice2018} are neutron-rich enough to produce a significant amount of $^{225}$Ac. This suggests that precise measurement of future kilonova late-time light curves could possibly be used to infer the presence of $^{225}$Ac.

\begin{figure}
\centering
\includegraphics[width=0.45\textwidth]{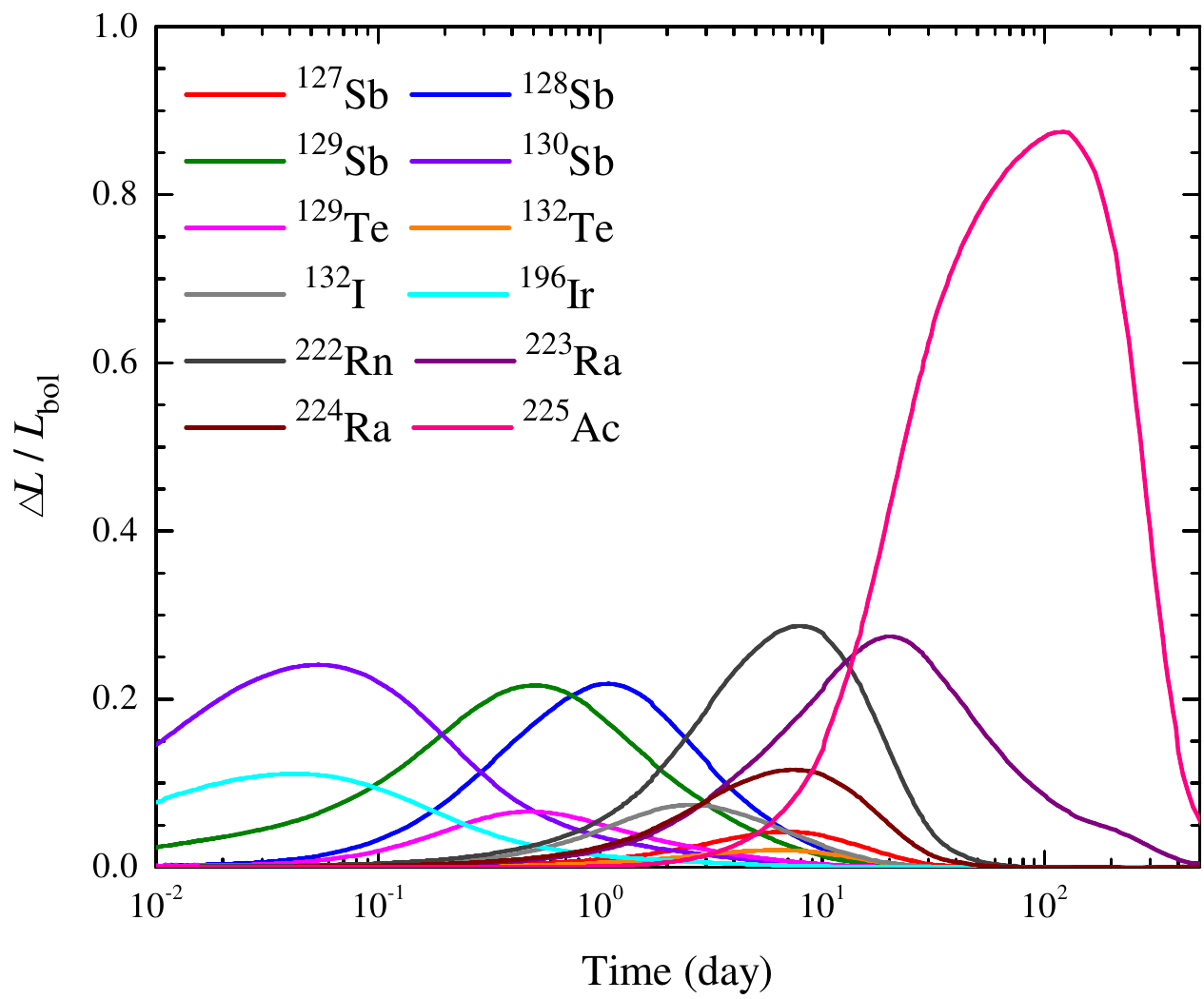}
\caption{The sensitivity of kilonova light curve to the dominant nuclides. Here $L_{\rm bol}$ represents the bolometric luminosity of kilonova emission, and $\Delta L$ represents the increase in luminosity caused by the doubling of the abundances of specific heavy elements.}
\label{dL}
\end{figure}

To explore the sensitivity of kilonova emission to the dominant nuclides, we calculate the ratio of luminosity increment ($\Delta L$) caused by the doubling the abundances of specific heavy elements, as shown in Figure~\ref{dL}. It is found that the early-time ($t\lesssim10$~days) kilonova light curve exhibits weak dependence on dominant nuclides, making it challenging to identify heavy elements contributing to the early-time light curve. However, at later times ($t\gtrsim20$~days), kilonova emission becomes inceasingly sensitive to $^{225}$Ac, indicating that the late-time kilonova light curve strongly depends on the yield of $^{225}$Ac. This feature can provide an important diagnostic for the nuclide composition in neutron star mergers. If the late-time observation up to 100~days after merger can be detected with sensitive instruments like the JWST, $^{225}$Ac could be distinguished from the bolometric light curve.

\begin{figure}
\centering
\includegraphics[width=0.45\textwidth]{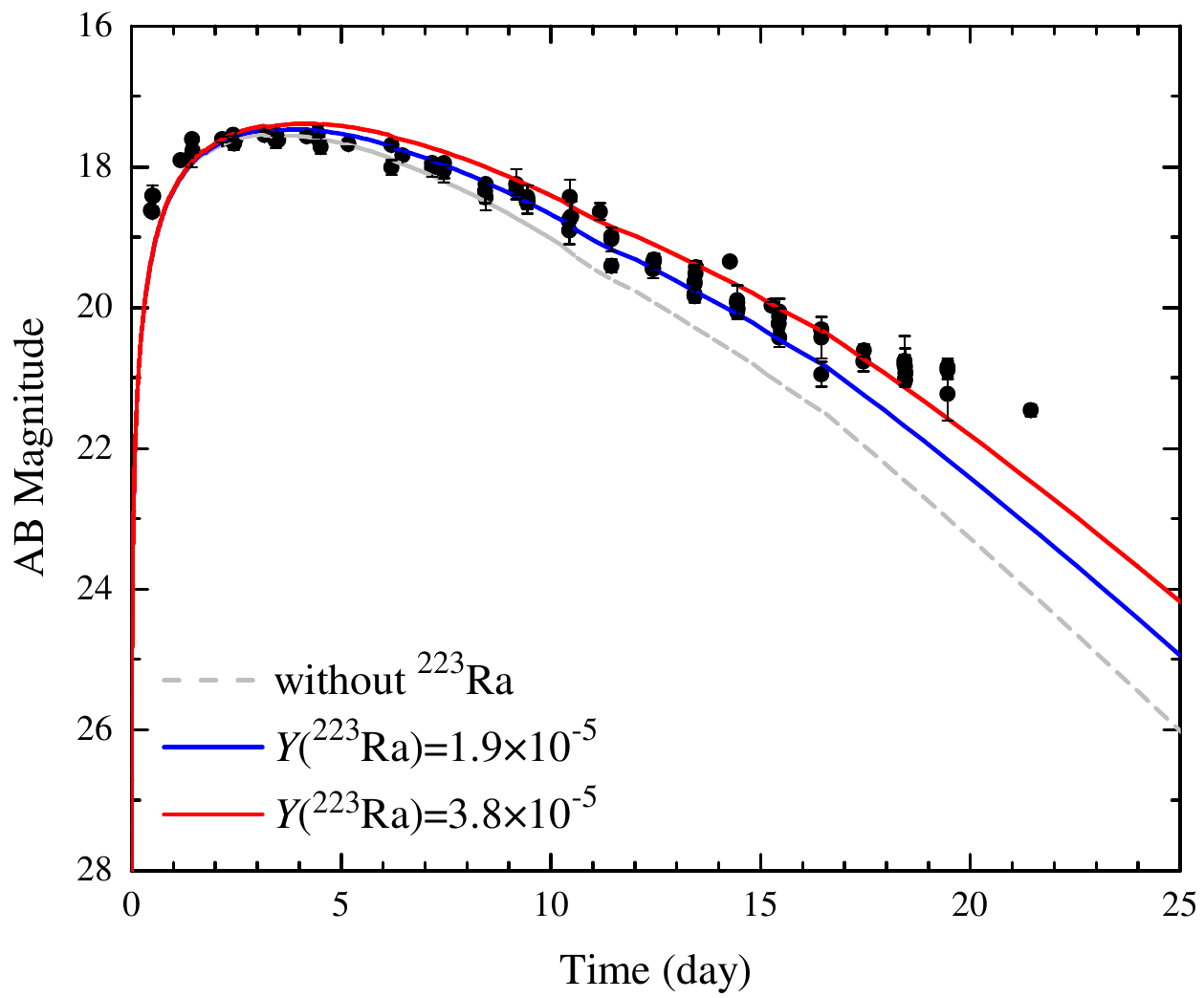}\\
\includegraphics[width=0.45\textwidth]{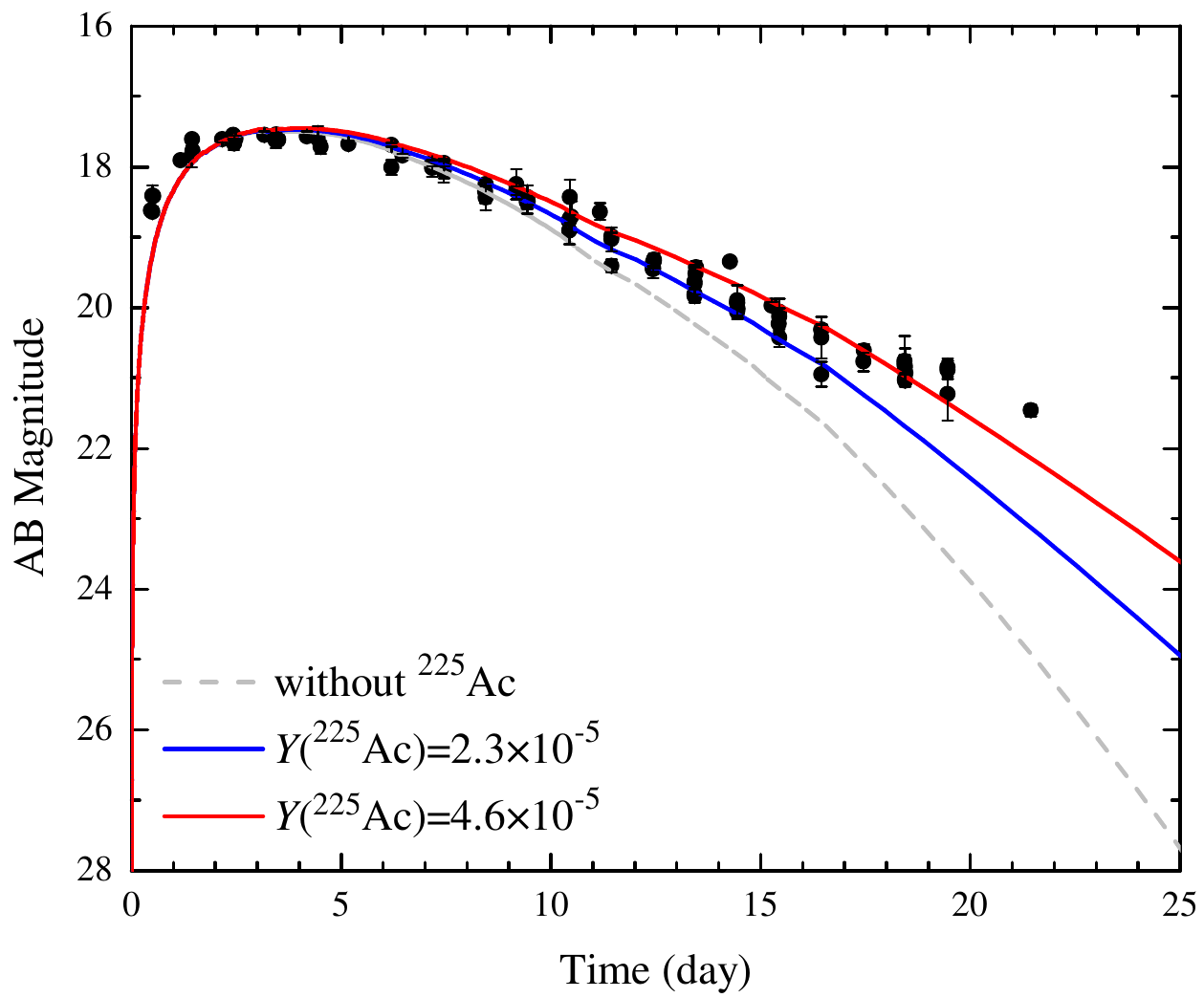}
\caption{K-band light curves with and without the contributions from $^{223}$Ra (top panel) or $^{225}$Ac (bottom panel) for a merger event assumed to be at a distance of $40$~Mpc, similar to that of GW170817/AT2017gfo. The observed data of AT2017gfo are obtainted from the combined data set \citep{Villar2017}. Blue solid lines represent the kilonova light curves obtained from $r$-process nucleosynthesis simulations, while red solid lines represent the light curves where the abundances of specific nuclides have been multiplied by a factor of 2.}
\label{Kband}
\end{figure}

The kilonova AT2017gfo provides a solid case for studying the nulide composition of merger ejecta, and observations in the near-infrared band have been obtained by the Gemini Observatory and the Hubble Space Telescope, as shown in Figure~\ref{Kband}. The near-infrared light curves can be fit by the kilonova model except for the data at $\sim20$~days (blue solid lines). As discussed earlier (see Figure~\ref{dL}), at around $20$~days, the kilonova emission is dominated by two $\alpha$-decay nuclei, $^{223}$Ra  (with an $\alpha$-decay half-life of $11.43$~days) and $^{225}$Ac. In simulations of $r$-process nucleosynthesis, the resulting abundances of $^{223}$Ra and $^{225}$Ac are $1.9\times10^{-5}$ and $2.3\times10^{-5}$, respectively. The gray dashed lines in Figure~\ref{Kband} show the scenario where there is no contribution from either $^{223}$Ra or $^{225}$Ac, resulting in the kilonova light curve in the near-infrared band dimming by almost one or two magnitudes at $\sim20$~days after the merger, respectively. To explain the late-time near-infrared data of the kilonova AT2017gfo, the required yield of $^{223}$Ra or $^{225}$Ac in the merger ejecta must be multiplied by at least a factor of $\sim2$ to produce the observed bolometric luminosity (red solid lines). This suggests that the late-time feature in the kilonova AT2017gfo at $\sim20$~days indicates the production of $^{223}$Ra or $^{225}$Ac at an abundance level of $\sim10^{-5}$, which is consistent with the results of \cite{Wu2019}. However, distinguishing the contribution of $^{225}$Ac from that of $^{223}$Ra is challenging due to their similar contributions at $\sim20$~days. Further observations in the near-infrared band are necessary to confirm the individual heavy elements responsible for the late-time kilonova light curves. Note that we employ the assumption of local thermal equilibrium in our analysis, which is reasonable for modeling kilonova light curve during the early stages. However, the kilonova emission transitions into a non-local thermal equilibrium in the later stages \citep{2022MNRAS.513.5174P,2023MNRAS.526.5220P,2023MNRAS.526L.155H}.

\begin{figure}
\centering
\includegraphics[width=0.45\textwidth]{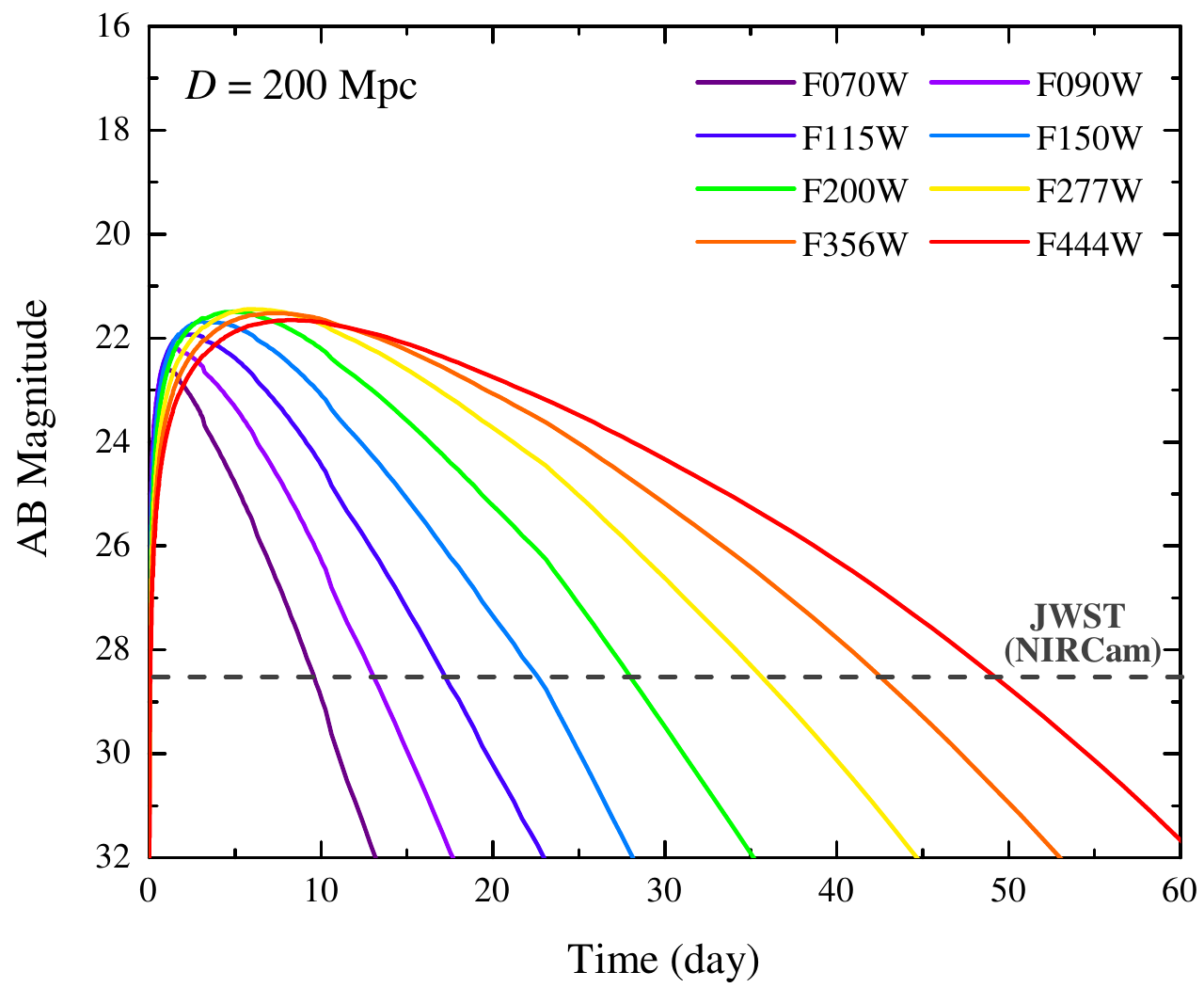}
\caption{Multi-band light curves for neutron star mergers. The ejecta mass is set to be $0.04M_{\odot}$ and the velocity is set to be $0.15c$, which are consistent with those inferred from the red kilonova component of AT2017gfo \citep{Kasen2017}. The merger is assumed to occur at a distance of 200~Mpc. The dashed lines represent the approximate observation limits of the NIRCam instruments with a $10^4$~s exposure time.}
\label{Magnitude}
\end{figure}

The JWST is expected to be a powerful instrument for detecting and studying kilonovae, particularly in the near-infrared band. In Figure~\ref{Magnitude}, we compare the resulting multi-band light curves to the sensitivity of JWST. Here the total ejecta mass is set to be $0.04M_{\odot}$ and the characteristic ejecta velocity is set to be $0.15c$, which are consistent with those inferred from the red kilonova component of AT2017gfo \citep{Kasen2017}. The merger is assumed to occur at a distance of $200$~Mpc, which represents the average reach of the Advanced LIGO/Virgo gravitational wave detectors during the fourth observing run. Adopting the limits of NIRCam's wide-band filters from the JWST website\footnote{https://jwst-docs.stsci.edu/jwst-near-infrared-camera/nircam-performance/nircam-sensitivity}, it is shown that kilonovae can be detectable by JWST up to $\sim50$~days. If a merger event occurs at a closer distance, such as the GW170817/AT2017gfo event at $40$~Mpc, JWST could potentially detect the corresponding kilonova up to $\sim100$~days. Therefore, the heavy element $^{225}$Ac that make a dominant contribution to the late-time kilonova can be distinguished in the future observations with JWST.

\begin{figure}
\centering
\includegraphics[width=0.45\textwidth]{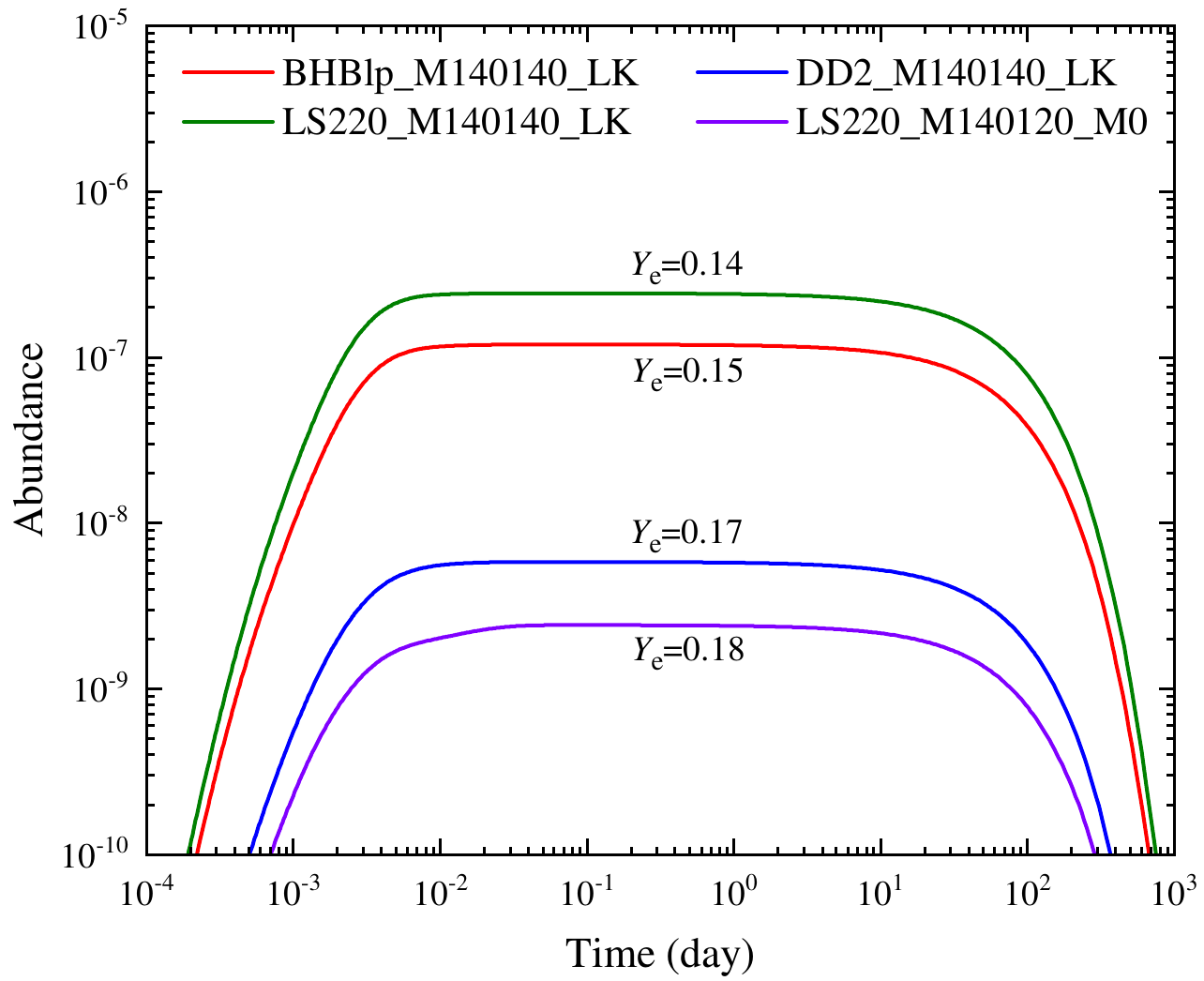}
\caption{Temporal evolution of the abundance of the spontaneous fission nucleus $^{254}$Cf.}
\label{Fission}
\end{figure}

We should emphasize that our calculations do not included the extra radioactive energy generated by fission fragments due to limited radiation data available for fission processes in the EDNF library. \cite{Zhu2018} suggested that the presence of the spontaneous fission nucleus $^{254}$Cf could potentially impact the late-time kilonova light curves, competing with the $\alpha$-decay of $^{225}$Ac. Figure~\ref{Fission} shows the temporal evolution of the abundance of spontaneous fission nucleus $^{254}$Cf. The resulting abundance of $^{254}$Cf in our $r$-process simulations spans the range of $\sim10^{-7}$ to $10^{-9}$. It is found that a lower electron fraction leads to a higher production of $^{254}$Cf. By taking into account the initial conditions of $r$-process simulations presented in \cite{Radice2018}, the abundance of $^{254}$Cf is consistently lower than the value ($\sim10^{-6}$) reported by \cite{Zhu2018} by at least one order of magnitude. This is reasonable because the initial conditions (electron fraction $Y_{\rm e} \simeq 0.15$) provided by \cite{Radice2018} are neutron-rich enough to generate actinides but not sufficient to produce a significant amount of $^{254}$Cf \citep{Wu2022}. Therefore, in our calculations, the contribution of the spontaneous fission nucleus $^{254}$Cf to the overall radioactive heating rates is relatively small compared to that of $\alpha$-decay chains. We also note that the choices of fission barriers and nuclear mass models may influence the production of $^{254}$Cf in $r$-process simulations \citep{Goriely2015,Barnes2016,Wanajo2018,Vassh2019,Zhu2021}.

\section{Conclusions and Discussions}
\label{conclusion}

Based on simulations of $r$-process nucleosynthesis, we have studied the energy fractions of dominant nuclides that contribute significantly to the energy generation of $\alpha$-particles, $\beta$-particles, and $\gamma$-rays. We have identified the heavy elements that dominate the kilonova emission within $\sim100$~days, including $^{127}$Sb, $^{128}$Sb, $^{129}$Sb, $^{130}$Sb, $^{129}$Te, and $^{132}$I. The $\alpha$-decay chains responsible for the late-time radioactive heating rate are also identified, including $^{222}$Rn, $^{223}$Ra, $^{224}$Ra, and $^{225}$Ac. These nuclide species could be distinguished from the kilonova light curves in future observation with sensitive instruments such as JWST. It is found that the late-time kilonova light curve ($t\gtrsim20$~days) is highly sensitive to the yield of $^{225}$Ac (with an $\alpha$-decay half-life of $10.0$~days, originating from the $\beta$-decay of $^{225}$Ra with a $\beta$-decay half-life of $14.9$~days), making it the most promising heavy elements to be identified in upcoming kilonova observations. Our results are consistent with previous studies \citep{Metzger2010,Lippuner2015,Wanajo2018,Wu2019}. We have also derived the characteristic time of dominant nuclides when they play a crucial role in the radioactive heating rate, which provides clues for identifying specific heavy elements in the late-time photometry.

The neutron star merger rate inferred from the LIGO/Virgo gravitational wave detectors is $R_{\rm BNS}=1540^{+3200}_{-1220}$~Gpc$^{-3}$~yr$^{-1}$ \citep{Abbott2017a}, and thus JWST is expected to detect $\sim52^{+107}_{-41}$ kilonova events per year following gravitational wave triggers. Recent results from the third observing run of LIGO/Virgo suggest a lower merger rate density of $R_{\rm BNS}=320^{+490}_{-240}$~Gpc$^{-3}$~yr$^{-1}$ \citep{Abbott2021}, which is broadly consistent with the predictions obtained from population synthesis models \citep{Belczynski2020}. Kilonovae are also expected to be discovered following short duration gamma-ray bursts (GRBs), which are commonly believed to originate from compact binary mergers \citep{Tanvir2013,Abbott2017c}. \cite{Rastinejad2021} have demonstrated that JWST is capable of detecting kilonova events following short GRBs at redshifts $z\lesssim1$, which increases the expected rate of detectable kilonovae.
The LIGO/Virgo network can provide a medium localization accuracy of 10~deg$^2$.To cover a localized sky area of 10~deg$^2$ uniformly, JWST would require $\sim50$~hours of total exposure time due to its limited field of view (9.7~arcmin$^2$). However, \cite{Bartos2016} proposed that the total search time could be reduced to 12.6~hours by focusing on known galaxies within 200~Mpc, since most neutron star mergers are expected to occur in or near galaxies. The search time could be further decreased by concentrating on galaxies with high star-formation rates that are more likely to host kilonovae. Therefore, the JWST will be able to rapidly locate kilonovae, enabling us to search for specific heavy elements in the early or late phases of the kilonova.
If the radioactive energy from fission fragments is included in the heating rate calculations, the resulting kilonova would be much brighter than the one without fission process. Consequently, it would be more easier to detect the kilonova in future observations.

\section*{Acknowledgements}

We thank Li-Xin Li for fruitful discussion and an anonymous referee for helpful comments.
This work was supported by the National Natural Science Foundation of China (Grant No.~12133003).

\begin{center}
{\bf ORCID iDs}
\end{center}

Meng-Hua Chen: https://orcid.org/0000-0001-8406-8683

En-Wei Liang: https://orcid.org/0000-0002-7044-733X

\section*{Data Availability}
 
The data underlying this paper can be shared on reasonable request to the corresponding author.



\bsp
\label{lastpage}
\end{document}